\documentclass[sigconf,screen=true]{acmart}

\acmSubmissionID{1658}

\usepackage{acronym}
\usepackage[skip=0pt]{caption}
\usepackage[inline]{enumitem}
\usepackage{booktabs}
\usepackage{makecell}
\usepackage{multirow}
\usepackage{subcaption}
\usepackage{geometry}
\usepackage{array}
\usepackage{xspace}

\AtBeginDocument{%
  \providecommand\BibTeX{{%
    \normalfont B\kern-0.5em{\scshape i\kern-0.25em b}\kern-0.8em\TeX}}}

\acrodef{TDS}{task oriented dialogue system}
\acrodef{AMT}{amazon mechanical turk}
\acrodef{ReDial}{recommendation dialogues}
\acrodef{LLM}{large language model}
\acrodef{IR}{information retrieval}
\acrodef{CRS}{conversational recommender system}
\acrodef{HIT}{human intelligence task}
\acrodef{ICC}{intraclass correlation coefficient}
\acrodef{KDE}{kernel density estimation}

\newcommand{\header}[1]{\vspace{1mm}\noindent\textbf{#1.}}

\allowdisplaybreaks

\newcommand{\halfnegskip}{\vspace*{-.625mm}}
\newcommand{\negskip}{\vspace*{-1.25mm}}
\newcommand{\dnegskip}{\vspace*{-2.5mm}}

\setcopyright{rightsretained}
\copyrightyear{2024}
\acmYear{2024}
\setcopyright{rightsretained}
\acmConference[SIGIR '24]{Proceedings of the 47th International ACM SIGIR Conference on Research and Development in Information Retrieval}{July 14--18, 2024}{Washington, DC, USA}
\acmBooktitle{Proceedings of the 47th International ACM SIGIR Conference on Research and Development in Information Retrieval (SIGIR '24), July 14--18, 2024, Washington, DC, USA}
\acmDOI{10.1145/3626772.3657712}
\acmISBN{979-8-4007-0431-4/24/07}



\author{Clemencia Siro}
\orcid{0000-0001-5301-4244}
\affiliation{%
  \institution{University of Amsterdam}
  \city{Amsterdam}
  \country{The Netherlands}}
\email{c.n.siro@uva.nl}

\author{Mohammad Aliannejadi}
\orcid{0000-0002-9447-4172}
\affiliation{%
  \institution{University of Amsterdam}
  \city{Amsterdam}
\country{The Netherlands}}
\email{m.aliannejadi@uva.nl}

\author{Maarten de Rijke}
\orcid{0000-0002-1086-0202}
\affiliation{%
  \institution{
  University of Amsterdam}
  \city{Amsterdam}
  \country{The Netherlands}}%
\email{m.derijke@uva.nl}

\renewcommand{\shortauthors}{Siro, Aliannejadi and de Rijke}

\begin{document}

\title[Rethinking the Evaluation of Dialogue Systems: Effects of User Feedback on Crowdworkers and LLMs]{Rethinking the Evaluation of Dialogue Systems:\\ Effects of User Feedback on Crowdworkers and LLMs}

\begin{abstract}

In ad-hoc retrieval, evaluation relies heavily on user actions, including implicit feedback. 
In a conversational setting such signals are usually unavailable due to the nature of the interactions, and, instead, the evaluation often relies on crowdsourced evaluation labels. 
The role of user feedback in annotators' assessment of turns in a conversational perception has been little studied.
We focus on how the evaluation of \acp{TDS}, is affected by considering user feedback, explicit or implicit, as provided through the follow-up utterance of a turn being evaluated. 
We explore and compare two methodologies for assessing \acp{TDS}: one includes the user's follow-up utterance and one without. We use both crowdworkers and \acp{LLM} as annotators to assess system responses across four aspects: relevance, usefulness, interestingness, and explanation quality.
Our findings indicate that there is a distinct difference in ratings assigned by both annotator groups in the two setups, indicating that user feedback does influence system evaluation. Workers are more susceptible to user feedback on usefulness and interestingness compared to \acp{LLM} on interestingness and relevance. User feedback leads to a more personalized assessment of usefulness by workers, aligning closely with the user's explicit feedback. Additionally, in cases of ambiguous or complex user requests, user feedback improves agreement among crowdworkers. These findings emphasize the significance of user feedback in refining system evaluations and suggest the potential for automated feedback integration in future research. We publicly release the annotated data.\footnote{ \url{https://github.com/Clemenciah/LLMCrowdDialogueEval/tree/main/Data}}

\end{abstract}

\maketitle

\acresetall

\section{Introduction}

Evaluation of systems has been an integral part of the \ac{IR} research agenda for decades~\citep[e.g.,][]{cleverdon-1966-factors}. Traditionally, IR evaluation has relied highly on user actions, including implicit feedback such as click-through rates. However, in a conversational setting such signals are not usually available due to the nature of the interactions. As a result, the evaluation of dialogue systems increasingly relies on human evaluation, leading to a growing interest in user-centric evaluation methods~\citep{Shuo-CRS}.
However, asking for explicit user feedback from a user can be intrusive and may negatively impact user experience~\citep{implicit-feedback}. Therefore, in recent years, the assessment of conversational systems has relied on crowdsourced evaluation, leveraging the collective wisdom of human annotators.

\header{Turn-level assessments} 
When gathering evaluation feedback on individual turns in a conversational interaction, various design methods have been considered in the past. 
These include deciding on the type of judgment scale, as well as formulating annotation guidelines and methods for presenting dialogues under assessment at the turn level~\citep{Roitero-finee-grained-relevance,siro2024context}. 
Recent strategies for presenting turn-level utterances to annotators involve two main approaches: one displays both the user's initial request and the system's response for evaluation~\citep{simulating-usat,Siro-UsatCRS}, and the other shows only the system's response~\citep{mehri-eskenazi-2020-usr}. 
The choice between these approaches often depends on the specific evaluation metric in question. 
The first method, similar to the query--document pair setting in ad-hoc retrieval systems, operates under the premise that the user's initial request offers sufficient context for annotators to make well-informed evaluations.

\begin{figure}[!b]
    \vspace*{1mm}
    \centering
    \includegraphics[width=\linewidth]{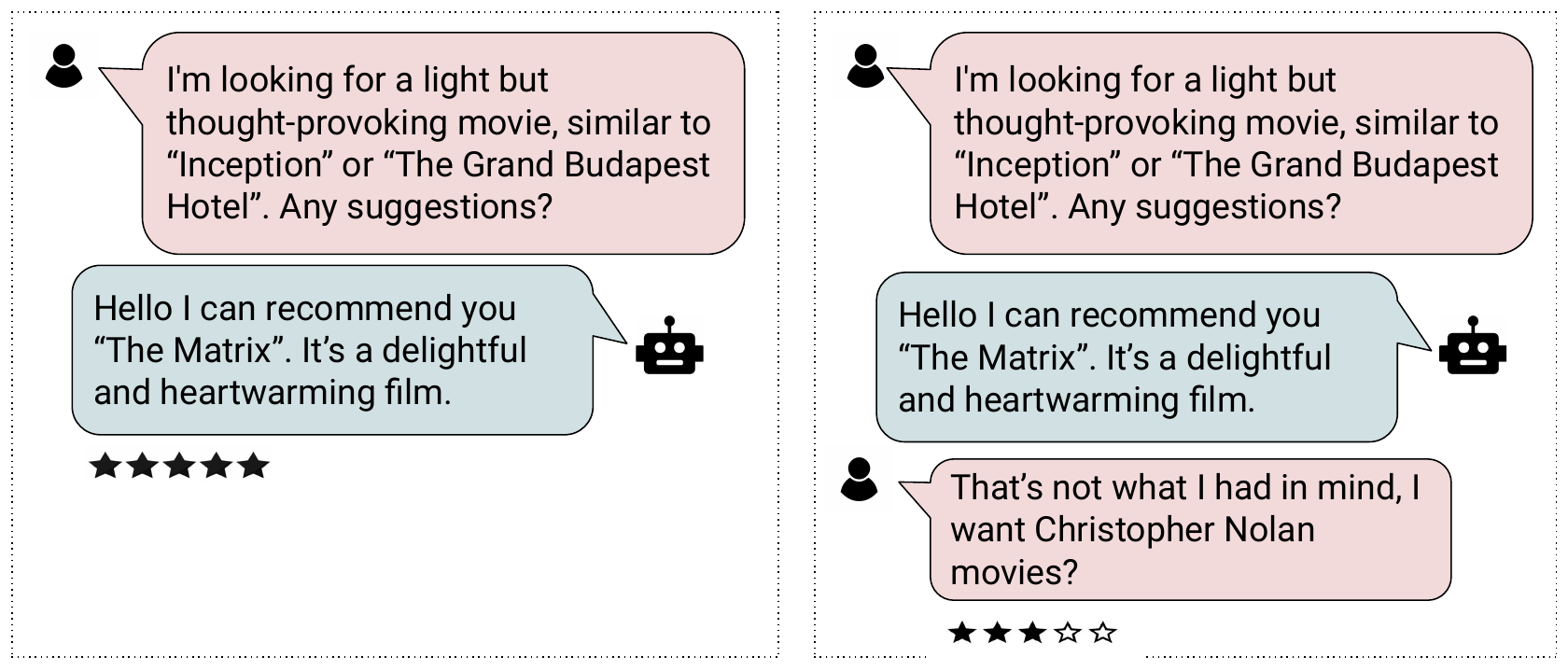}
    \caption{A dialogue showing an example of a complex user request with (right) and without (left) the user feedback. The star ratings show the assessment of external assessors judging the usefulness of the system utterance. As can be seen, based on the follow-up utterance the assessors lower their usefulness rating aligning with the user feedback.}
    \label{fig:dial-example}
     \vspace*{-.5mm}
\end{figure}

\header{Follow-up utterances}
Users often do not articulate all of their intentions in a single request. 
Rather, they engage in an iterative dialogue, clarifying and refining their intentions through successive exchanges~\citep{AliannejadiZCC19}. 
Their queries can be multifaceted, ambiguous, or overly generic, which complicates the process of evaluating individual turns. 
E.g., in Fig.~\ref{fig:dial-example}, the user poses a multifaceted query, leaving substantial room for interpretation. 
First, the user is looking for a movie suitable for an evening watch, which typically suggests something not overly long or intense. Second, they desire a film that is ``light but thought-provoking,'' implying a blend of easy-to-digest content with depth in storytelling. 
Last, by referencing specific movies like ``Inception'' and ``The Grand Budapest Hotel,'' the user indicates a preference for a certain style or genre --- perhaps celebratory narratives with unique storytelling or visually engaging films. 
Annotators tasked with evaluating the system's response to this query must consider these multiple layers.

When annotating turns in a conversational interaction, the complexity for annotators lies in assessing the system's response not just for its relevance to an overt request, e.g., for a movie recommendation, but also for its alignment with nuanced, implicit preferences indicated by the user. 
Systems may not always successfully address all aspects of such requests. 
When the system's response only partially meets the query's criteria, it becomes challenging for annotators to gauge which aspect of the request was most critical to the user. 
We believe that this is where a user's follow-up utterance be particularly informative. 
A user's subsequent response may provide valuable insights into what they valued most in their original request, giving annotators a clearer indication of the user's priorities. 
Thus, follow-up utterances may serve as crucial cues, helping annotators make a more informed assessment of the system's performance, particularly in how well it navigates and prioritizes the multifaceted aspects of a user's complex request.

\textit{Is a user's follow-up utterance crucial in ensuring evaluations align with actual user needs, especially since annotators, as external evaluators, may not fully grasp the user's perspective or context? }
We hypothesize that annotators who have access to a follow-up utterance produce more accurate and user-centric evaluations, improving the quality of evaluation labels in the process.

\header{Research goals}
We investigate the effect of a user's follow-up utterance on the annotation of turns in a \ac{TDS}. 
We conduct experiments with two types of annotators: human and \ac{LLM}-based.
Both types of annotators are asked to provide annotations of turn-level system responses along four dimensions:
\emph{relevance}, \emph{usefulness}, \emph{interestingness}, and \emph{explanation quality}, on a 100 level scale, following~\citep{Roitero-finee-grained-relevance}.
We consider two contrastive setups for annotators to provide these annotations:
\begin{enumerate*}
\item [(\ref{SetupOne})] does not consider the user's follow-up utterance, and
\item [(\ref{SetupTwo})] does consider the user's follow-up utterance.
\end{enumerate*}
In addition, we collect data on what sources of information human annotators rely on to arrive at their judgments. With this data, we aim to examine the bias introduced by these sources to the crowdsourced labels. 

We use a subset of the \ac{ReDial}~\citep{li-redial} dataset to address the following research questions:
\begin{enumerate*}[label=(\textbf{RQ\arabic*})]
    \item How does a user's implicit feedback from the follow-up utterance influence the evaluation labels collected from both human annotators and \acp{LLM}? \label{item:RQ1}
    
    \item When is implicit user feedback significant in the evaluation of \acp{TDS}? \label{item:RQ2} \looseness=-1

    \item What are the annotators' perceptions in terms of the sources of information they rely on to make their assessments, and what are the potential biases might that have on their performance? \label{item:RQ3} \looseness=-1
\end{enumerate*}

\header{Findings}
Our findings indicate that both the crowdworkers and the \ac{LLM} exhibit sensitivity to user cues from follow-up utterances. There is a significant difference in the mean ratings from both annotators except for relevance when follow-up utterance is included, indicating user feedback does influence system evaluation. Workers are more susceptible to user feedback in usefulness and interestingness, compared to \acp{LLM} in interestingness and relevance. Specifically, there is a clear distinction in relevance and usefulness ratings of crowdworkers \ref{SetupTwo} ratings unlike in \ref{SetupOne} where these aspects are often conflated. This indicates that crowdworkers not only evaluate response usefulness based on topical relevance but also align with user needs and preferences expressed in follow-up utterances. This suggests that follow-up utterances enable a more personalized assessment of usefulness, aligning closely with the user's explicit feedback. 
In \ref{SetupTwo}, we observe an increase in annotator agreement. 
This is particularly evident in scenarios characterized by uncertainty in user requests. 
Complex user requests with multiple criteria or preferences posed challenges during evaluation, but follow-up utterances helped to clarify the user intent. 
Similarly, generic user requests, initially broad and challenging to address, became more focused on follow-up utterances, allowing human-/LLM-based annotators to tailor their assessment effectively.

These findings not only show the significance of user feedback in system evaluation but also provide a foundation for integrating user feedback in the automatic evaluation of conversational systems.

\dnegskip
\section{Related Work}
\halfnegskip

Recent studies emphasize evaluating \acp{TDS} through a user experience lens~\citep{Deriu2020SurveyOE}. Traditionally, \acp{TDS} were assessed primarily for task completion. While task completion remains a fundamental criterion, there is a growing recognition that solely measuring task success may not provide a comprehensive understanding of system performance~\citep{Siro-UsatCRS}. Consequently, there is a shift towards incorporating user experience metrics into the evaluation of \acp{TDS}~\citep{Siro-satTDS}. Here, we provide a brief overview of the studies on \ac{TDS} evaluation from the perspective of user feedback, evaluation bias, and LLMs. \looseness=-1

\vspace{-3mm}

\subsection{User feedback}
In web search, implicit user signals including click-through rates and dwell time on search results are available in vast amounts and these signals are leveraged to evaluate a system and continually improve the search results~\citep{Kim-insitu,implicit-feedback,Yiqun-mousemovements,searchsuccess,Ryen-clicks}.
However, such signals are not accessible for conversational systems due to their interactive nature. Consequently, automatic evaluation of conversational systems would primarily rely on explicit user feedback in the form of ratings~\citep{Praveen-joint,Jason-onlineofflineusat}, which could be intrusive and lead to poor user experience~\citep{Praveen-multidomain}. In this work, we propose using the user's next utterance as a proxy for both explicit and implicit user feedback.  For instance, when a user expresses satisfaction or dissatisfaction in their next utterance following a system recommendation, it represents explicit feedback that should not be disregarded when assessing system performance.
Our study investigates how user feedback from the next utterance influences the evaluation labels provided by both human- and LLM-based annotators.

\vspace{-2.5mm}
\subsection{Bias in crowdsourcing evaluation labels}
\vspace{-1mm}
The use of crowdsourcing for evaluating \acp{TDS} in \ac{IR} research, while offering scalability and diversity, brings inherent biases. Research, such as \citep{Carsten-cognitive-bias,DBLP:cognitiveload-ds,Leif-Cognitivebias}, highlights cognitive biases and load as significant factors influencing crowdworker judgments, which can skew the evaluations. For example, workers' preconceived notions or mental fatigue might lead to inconsistent results. Further, \citet{mitigating-workerbias} and \citet{crowdworker-relevance-strategy} emphasize the impact of workers' relevance strategies and personal biases on their assessment of \ac{IR} systems. To mitigate these biases, strategies such as task design adjustments and worker training are suggested~\citep{Gadiraju-taskclarity}. For instance, presenting tasks neutrally and providing clear, unbiased instructions can help reduce bias, as discussed in \citep{mitigating-workerbias}. Additionally, the choice and implementation of judgment scales, as explored in \citep{judgmentscale-assessorbias}, play a crucial role in assessor bias. Different from previous studies, in this work we focus on assessing how the sources of information relied on by workers to make their judgment bias their evaluation labels. \looseness=-1

\vspace{-2.5mm}
\subsection{LLMs as annotators}
\vspace{-1mm}
Recently, there has been a notable surge in the use of \acp{LLM} as annotators in various tasks~\citep{llm-survey-chang}. These models show good performance comparable to human annotators and in some cases outperform them~\citep{Chatgpr-outperforms-workers}. Additionally, they have proven to reduce the time and cost for annotation, making them a preferred choice compared to human annotators~\citep{annotationcost-reductionwithgpt}. However, most research efforts have primarily focused on assessing how well \acp{LLM}' labels correlate to human labels~\citep{GPT-goodannotator,AnnoLLM}. There has been a relative lack of investigation into whether \acp{LLM} are susceptible to the same influencing factors as crowdworkers.

Several studies have delved into understanding the factors that impact crowdworkers, encompassing aspects like task design, judgment scales, and protocols designed to enhance the quality of annotations~\citep{DBLP:conf/sigir/RoiteroSFSMD20,Kazai-humanfactors,taskabandonement}. These studies have contributed valuable insights into optimizing crowdworker performance. However, a similar examination of the impact of these factors on \acp{LLM} is notably absent from the research landscape. We contribute towards understanding how task design influences evaluation labels assigned by \acp{LLM}. 
We investigate the influence of user feedback from the user's follow-up utterance on the evaluation of \acp{TDS} by both crowdworkers and \acp{LLM}.

\dnegskip
\section{The Annotation Task}
\halfnegskip
Our objective is to understand the influence of user feedback from the user's follow-up utterance on the evaluation of \acp{TDS}. 
We conduct our study as an annotation effort with crowdworkers from \ac{AMT}~\citep{mturk}. 
Additionally, due to the increased use of \acp{LLM} as annotators~\citep{AnnoLLM,GPT-goodannotator} we seek to understand how \acp{LLM} are affected by feedback from the user's next utterance. 
User feedback in this case can be either implicit or explicit. 
Explicit feedback refers to straightforward, direct responses from users, like specific comments on a certain dialogue aspect (e.g., ``I don't like this movie''). 
Implicit feedback is more subtle, encompassing aspects like tone or contextual hints within the user's follow-up utterance (e.g., ``Thanks for the suggestion. How about some action movies?''). 
We gather annotations for four fine-grained dialogue qualities: relevance, usefulness, interestingness, and explanation quality, across two experimental conditions.

\vspace{-2.5mm}
\subsection{Dialogue qualities}
\vspace{-1mm}
We experiment with four dialogue qualities in the domain of \acp{TDS}~\citep{explanationquality-Balog,Siro-UsatCRS} that have been investigated extensively, elaborated below. 

\noindent
\textbf{Relevance.}
The relevance~\citep{Jiang-insitu-relevance,Maddalena-relevanceME,Alonso-relevance-crowdsourcing,Siro-UsatCRS} of a dialogue response is a crucial factor in assessing the effectiveness of a \ac{TDS}. To evaluate relevance, we ask the workers to determine how well the system's response addresses the user's request. This aspect gauges the system's ability to understand and appropriately respond to user input. \looseness=-1

\noindent
\textbf{Usefulness.}
Usefulness~\citep{Usefulness-searchresults,Mao-usefulness-mobilesearch,siro2024context} of a dialogue response pertains to its practical value from the user's perspective. Apart from just being relevant, workers assess whether the system's response gives additional information to the user on the recommended item. In \ref{SetupTwo} the user's perspective is captured by asking workers to rely on the user's follow-up utterance to gauge the usefulness of the system response. E.g., if a user says, ``I have already watched that,'' it suggests that the recommendation is not new or helpful to the user, even though it might be relevant. Usefulness helps to measure the system's overall utility in real-world scenarios.

\noindent
\textbf{Explanation quality.}
Understanding how well a \ac{TDS} communicates its reasoning is important for user trust and comprehension. Explainability in \ac{IR} systems has witnessed a notable surge in recent times~\citep{explainableCRS-Shuyu,explanationquality-Balog,Ma-explainable-recommendation,Explainable-rec-survey}. 
Following~\citet{explainableCRS-Shuyu} we instruct workers to assess explanation quality, by evaluating the clarity and informativeness of the system's justifications or explanations accompanying its responses. This aspect provides insights into the system's transparency and user-friendly communication.

\noindent
\textbf{Interestingness.}
Beyond system functionality, the interestingness~\citep{Siro-satTDS} of a system response adds a subjective layer to the evaluation. Workers are asked to evaluate whether the system's responses are engaging, captivating, or exhibit qualities that make the interaction more enjoyable for the user. This encapsulates the language used to make recommendations by the system. This aspect contributes to a holistic assessment of user experience.

\vspace{-2.5mm}
\subsection{Data}
\vspace{-1mm}
We use the \ac{ReDial} dataset~\citep{li-redial}, a well-known collection of over 11,000 dialogues specifically focused on movie recommendations. We sample the dialogue turns for annotation by focusing on selecting user utterances that explicitly request movie recommendations or express movie preferences. Phrases like ``I prefer,'' ``recommend me,'' and ``my favorite'' were key indicators in this selection process. This approach ensured that our sampled data contained user utterances that were explicit and straightforward, facilitating a more accurate assessment of the dialogue system's responses.

Similar to~\citet{explainableCRS-Shuyu}, we observe a lack of in-depth explanations in the \ac{ReDial} dialogues. Our initial analysis of the dataset indicates that longer system utterances often include attempts to explain movie recommendations, whereas shorter ones do not. As a result, we selected system utterances with more than 14 words~(average length of system responses in the dataset)  to better focus on responses that are more likely to include explanations.
In total, we sampled 100 unique dialogue turns from the dataset, each representing a different conversation.

\vspace{-2.5mm}
\subsection{Annotation scale}
\vspace{-1mm}
Following the approach outlined in \citep{Roitero-finee-grained-relevance}, our study adopts the S100 scale for evaluation purposes. This scale is employed through a sliding window mechanism, allowing annotators to provide detailed feedback on the dialogue systems. The sliding scale's interactive nature enables a more precise and flexible assessment compared to traditional binary or categorical scales. To enhance its usability and ensure intuitive responses, the default value on this slider is set to 0. This design choice is based on the rationale that a neutral starting point encourages annotators to consciously adjust the slider based on their judgment of the dialogue turn's effectiveness, rather than being biased by any preset values. 
To ensure consistency and accuracy in evaluations, we provide the annotators with several examples demonstrating how to effectively use the S100 scale.
We adopt the same scale for annotation with \ac{LLM}.

\vspace{-2.5mm}
\subsection{Preliminary experiments}
\vspace{-1mm}
Our research included preliminary experiments to refine the design and methodology. These experiments assessed the practicality of our setups, refined annotation guidelines, and identified data collection challenges. Two setups were tested: 

\begin{enumerate}[leftmargin=*,nosep,label=\textbf{Exp~\arabic*}]
    \item  \emph{Single worker, two conditions}: Workers evaluated a dialogue turn under two conditions within a single \ac{HIT}. The only difference was the presence or absence of the user's follow-up utterance, which served as user feedback. \label{exp1}
    
    \item \emph{Random assignment of conditions}: Workers were randomly assigned to one of the two conditions to incorporate diverse perspectives and reduce potential biases, gathering a range of rationales behind annotator evaluations. \label{exp2}
\end{enumerate}

\header{Preliminary results}
We used 13 dialogue turns, primarily focusing on comparing \ref{exp1} and \ref{exp2} to determine the most effective approach. The mean ratings obtained from both setups indicated a high degree of consistency in annotator assessments for relevance and usefulness, suggesting that both methods performed similarly in these aspects. However, an interesting observation emerged concerning annotation time and the diversity of justifications. In \ref{exp1}, resulted in shorter annotation times but exhibited limited diversity in justifications. In contrast, \ref{exp2}, yielded a more diverse set of justifications. Considering these findings, we decided to proceed with the \ref{exp2} setup for our main experiments. This choice was motivated by the goal of obtaining a broader and more diverse range of annotations and justifications, a critical requirement for the comprehensive evaluation of dialogue systems in our study.

\vspace{-2.5mm}
\subsection{Experimental conditions}
\vspace{-1mm}
Following \ref{exp2}, we designed two distinct experimental conditions to evaluate the effect of user feedback on the evaluation of \acp{TDS} with human annotators, as well as \acp{LLM}.

\begin{enumerate}[label=\textbf{Setup~\arabic*},leftmargin=*,nosep]    
    \item Following the conventional annotation method, this condition provides only the initial user query and the system's response to the annotators and \acp{LLM}, omitting the user's follow-up utterance. This setup focuses on evaluating the \ac{TDS} based on a single interaction, reflecting the traditional approach in dialogue annotation. \label{SetupOne}

    \item This condition incorporates the user's follow-up utterance along with the initial query and the system's response. The aim is to allow annotators and \acp{LLM} to evaluate the \ac{TDS} within the full context of the conversation, assessing the impact of subsequent user feedback on annotations. \label{SetupTwo}
\end{enumerate}

\vspace{-2.5mm}
\subsection{Human annotators}
\vspace{-1mm}
For this task, we recruited master workers from \ac{AMT}. We employed multiple \ac{HIT} templates to conduct our study, aiming to investigate the impact of the user's next utterance on annotator ratings for various aspects. We collected annotation labels for relevance, usefulness, interestingness, and explanation quality in the two experimental conditions. Each aspect was annotated in a separate \ac{HIT}. Importantly, we did not disclose the research angle to the annotators, framing it as an annotation effort.

During each \ac{HIT}, we provided the annotators with instructions, definitions of the aspect to be assessed, and examples. We maintained consistent instructions across all aspects and setups, with variations limited to definitions and examples. In each \ac{HIT}, annotators rated the aspect and provided justifications for their ratings, a practice known to reduce randomness and enhance assessment quality~\citep{rationales}. Additionally, we sought to understand the sources of information relied on by annotators to make their assessments, asking them to select sources they considered when making their assessments, including personal knowledge, external sources such as web searches, educated guesses, user's request, system response, user's feedback, and an ``other'' option for sources not covered in the provided options.

84 unique workers participated in the study (46 female and 38 male), with an average age of 30--45. Each worker received a reward of \$0.4 per \ac{HIT}, which was determined based on the minimum wage.

\vspace{-2.5mm}
\subsection{\ac{LLM} as annotator}
\vspace{-1mm}
Since \acp{LLM} have shown a notable performance as annotators, we investigate whether \acp{LLM} are influenced to a comparable degree as human annotators with user feedback from the user's follow-up utterance. This investigation seeks to shed light on the potential similarities and differences in how \acp{LLM} and human annotators respond to such contextual input, ultimately contributing to an understanding of \ac{LLM} behavior in annotation tasks.

We used ChatGPT~(gpt-3.5-turbo API\footnote{Temperature = 1, Top p = 1}) a subseries of GPT models~\citep{Brown-gpt} for the annotation task. 
Typically, crowdworkers are provided with annotation examples to improve the quality of the labels they assign, similarly, we provide the same to ChatGPT, thus using it in the few-shot setting. To ensure consistency, we replicate the experimental conditions used for human annotators across the four aspects. In our experiments, we employ two distinct experimental conditions, varying the prompts and data presentation to align with these setups.\footnote{The prompts are available in our \nameref{sec:appendix} in Sec.~A.
}
For each aspect, we provided ChatGPT with the corresponding human annotation instructions. This approach allows us to comprehensively assess the performance of ChatGPT in generating evaluation labels in a manner that mirrors human annotation practices.

\dnegskip
\section{Crowdsourced Judgments} \label{judgments}
\halfnegskip
Before addressing our research questions (in Sec.~\ref{section:RQ1}--\ref{section:RQ3}), we examine the judgments collected through crowdsourcing.

\header{Internal agreement}
To assess the quality of labels collected from the crowdworkers, we compute pairwise Cohen's Kappa and report the results in Tab.~\ref{tab:icc-values}. The Kappa scores indicate varying levels of agreement in different evaluation setups and aspects: 

\textit{Relevance} shows a substantial agreement in \ref{SetupOne}, compared to \ref{SetupTwo}~(Kappa: 0.69 vs.\ 0.57). This indicates that the inclusion of the user's follow-up utterance during evaluation introduces complexity, impacting crowdworkers' judgments. 
Both setups in \textit{usefulness} exhibit high agreement, indicating that the presence or absence of user follow-up utterances has minimal influence on crowdworkers' perceptions of the system's response utility. This suggests that, overall crowdworkers are consistent in the evaluation of usefulness. 
\ref{SetupOne} shows a moderate agreement (Kappa: 0.28) in evaluating \textit{interestingness} while \ref{SetupTwo} has a slightly lower agreement (Kappa: 0.22), reflecting the added complexity introduced by the user's follow-up utterance. 
In \emph{explanation quality}, \ref{SetupTwo} exhibits higher agreement (Kappa: 0.28 vs.\ 0.23), possibly because it allows crowdworkers to assess explanations within a broader context that includes both the initial response and the user reaction in the follow-up. 

We make similar observations with the \ac{ICC} values in the two setups. In general, workers exhibit a substantial to moderate agreement for all aspects. The low agreement in interestingness and explanation quality can be attributed to the nature of their subjectivity and the large scale of evaluation.

\begin{table}[]
    \centering
    \caption{\Ac{ICC} and pair\-wise Cohen's Kappa for all aspects across both set\-ups.}
    \begin{tabular}{l cc cc }
    \toprule
    & \multicolumn{2}{c}{ICC} & \multicolumn{2}{c}{Kappa}\\
    \cmidrule(r){2-3}
    \cmidrule(r){4-5}
        Aspects & Setup~1 & Setup~2 & Setup~1 & Setup~2  \\
        \midrule
        Relevance & 0.8358 & 0.7427 & 0.6978 & 0.5701  \\
        Usefulness & 0.7553 & 0.7419 & 0.5892 & 0.5631\\
        Interestingness & 0.5456 & 0.4685 & 0.2825 & 0.2231  \\
        Explanation quality & 0.5136 & 0.5351 & 0.2380 & 0.2812 \\
        \bottomrule
    \end{tabular}
    
    \label{tab:icc-values}
\end{table}

\begin{figure}[!t]
    \centering
    \begin{subfigure}{0.4\columnwidth}
        \includegraphics[width=\linewidth]{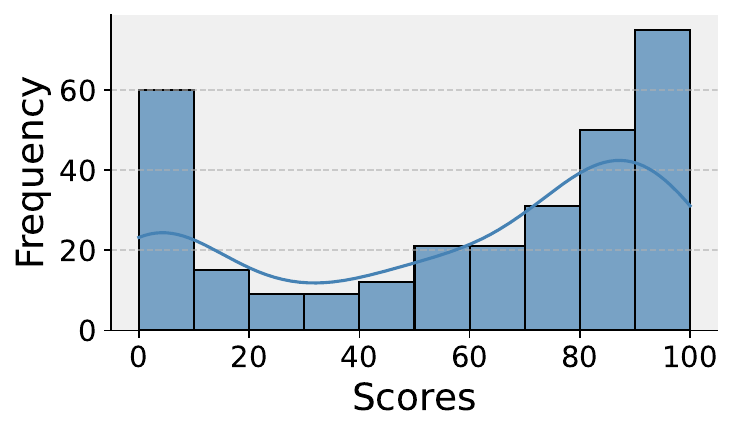}
        \vspace*{-6mm}
        \caption{Relevance}
        \label{fig:relworker-dist-s1}
    \end{subfigure}
    \quad
    \begin{subfigure}{0.4\columnwidth}
        \includegraphics[width=\linewidth]{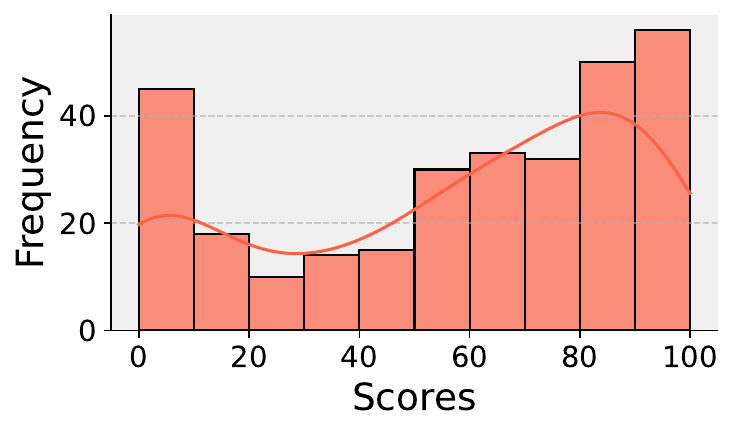}
        \vspace*{-6mm}
        \caption{Relevance}
        \label{fig:relworker-dist-s2}
    \end{subfigure}  
    \\
    \begin{subfigure}{0.4\columnwidth}
        \includegraphics[width=\linewidth]{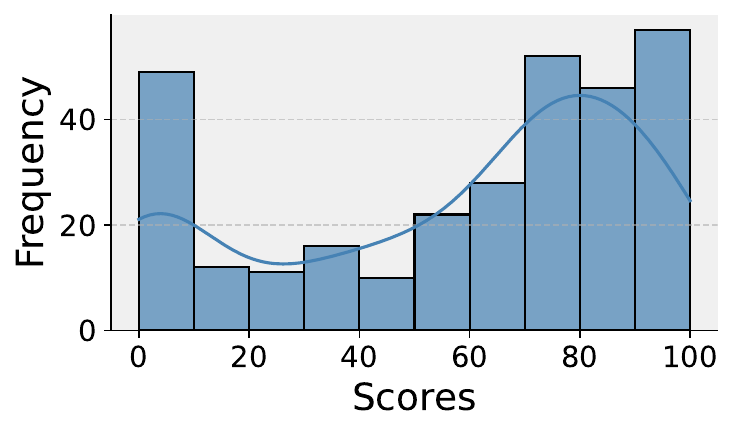}
        \vspace*{-6mm}
        \caption{Usefulness}
        \label{fig:useworker-dist-s1}
    \end{subfigure}
    \quad
    \begin{subfigure}{0.4\columnwidth}
        \includegraphics[width=\linewidth]{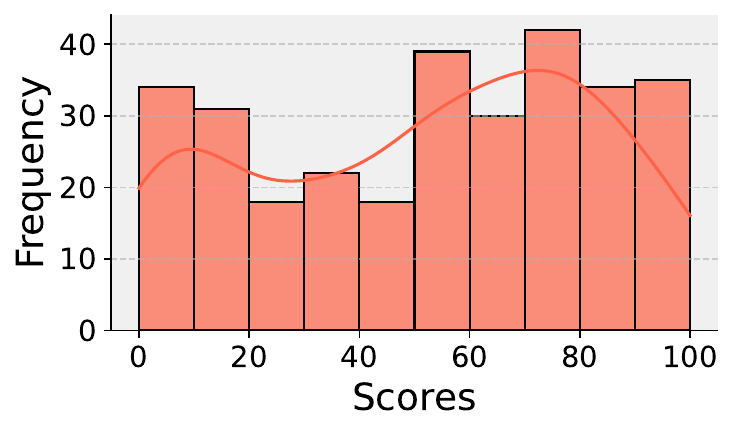}
        \vspace*{-6mm}
        \caption{Usefulness}
        \label{fig:useworker-dist-s2}
    \end{subfigure}  
    \\
    \begin{subfigure}{0.4\columnwidth}
        \includegraphics[width=\linewidth]{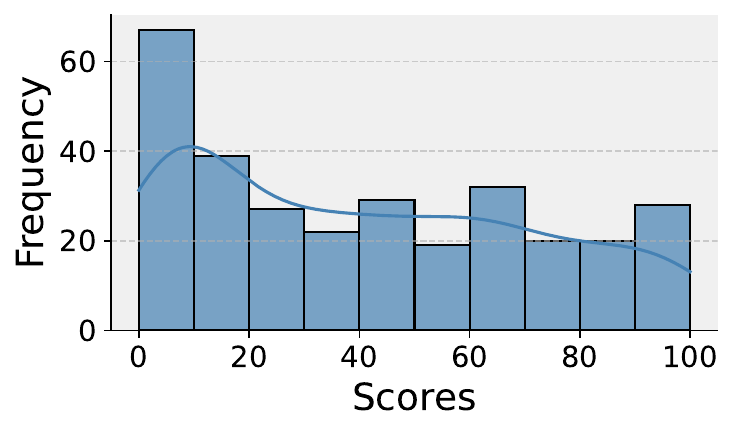}
        \vspace*{-6mm}
        \caption{Interestingness}
        \label{fig:intworker-dist-s1}
    \end{subfigure}
    \quad
    \begin{subfigure}{0.4\columnwidth}
        \includegraphics[width=\linewidth]{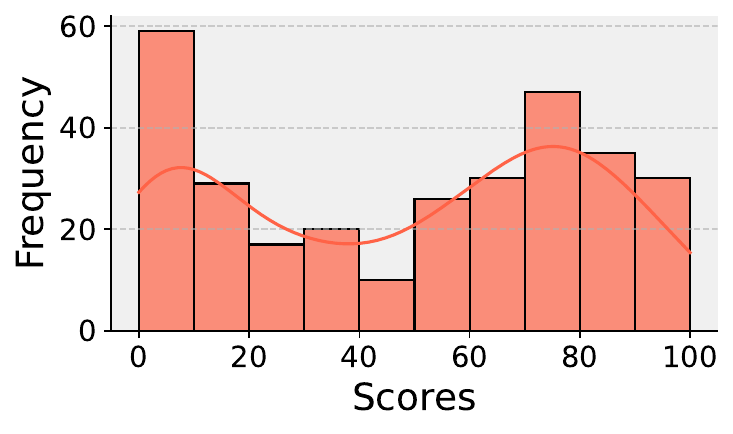}
        \vspace*{-6mm}
        \caption{Interestingness}
        \label{fig:intworker-dist-s2}
    \end{subfigure}  
    \\    
    \begin{subfigure}{0.4\columnwidth}
        \includegraphics[width=\linewidth]{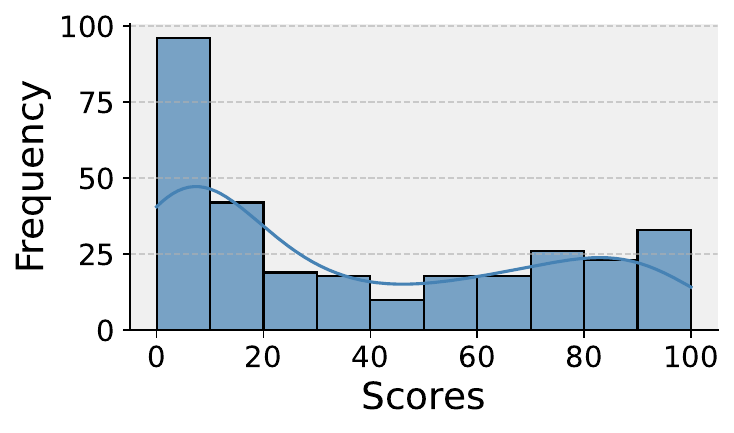}
        \vspace*{-6mm}
        \caption{Explanation quality}
        \label{fig:EQworker-dist-s1}
    \end{subfigure}
    \quad
    \begin{subfigure}{0.4\columnwidth}
        \includegraphics[width=\linewidth]{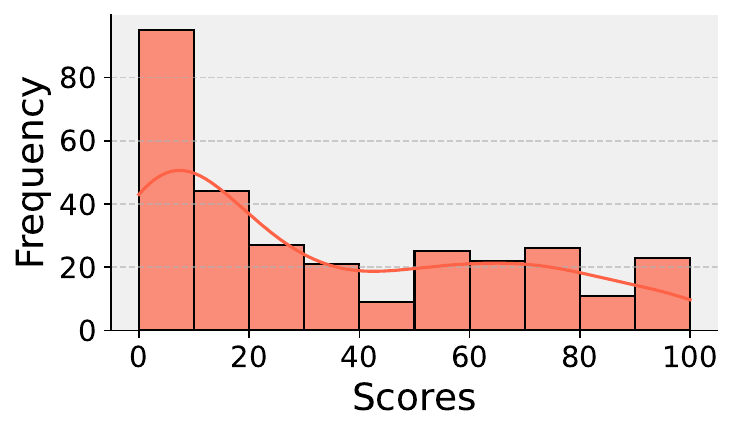}
        \vspace*{-6mm}
        \caption{Explanation quality}
        \label{fig:EQworker-dist-s2}
    \end{subfigure}  
    \\        
    \caption{A comparison of individual worker scores distributions for \ref{SetupOne}~(left column) and \ref{SetupTwo}~(right column). }
    \label{fig:worker-distributions}
\end{figure}

\header{Crowdworker judgments}
Fig.~\ref{fig:worker-distributions} shows the distributions of scores provided by crowdworkers for the four dialogue qualities over the two setups.  
In line with the findings of \citet{Roitero-finee-grained-relevance}, we see that 60\% of the scores are multiples of 5 or 10, showing that crowdworkers tend to select such values as their judgments in an S100 range. \looseness=-1

\emph{Relevance scores} in both setups display a long-tailed distribution toward the extremes, as shown in Fig.~\ref{fig:relworker-dist-s1} and \ref{fig:relworker-dist-s2}.  The median rating in \ref{SetupOne} is 71 with 65 for \ref{SetupTwo}, suggesting that including the user's follow-up utterance leads to lower relevance scores from the workers. The scores for both setups range between 0 to 100.

In \ref{SetupOne}, there is a noticeable drop in \textit{usefulness scores} within the 20--40 range, as seen in Fig.~\ref{fig:useworker-dist-s1}. This setup also indicates similarity in the distribution of relevance and usefulness scores, suggesting that, without user feedback, workers tend to rate usefulness similarly to relevance due to limited information. In \ref{SetupTwo}, which includes user follow-up, there is a decrease in responses rated as not useful (0--20) and an increase in the 30--70 range. This indicates that some responses are considered useful even when not directly relevant to the user's initial query, possibly because users found the recommended movie intriguing or had not considered it before. The median scores are 70 and 56 respectively in Setups 1 and 2.

 The \emph{interestingness} aspect is highly subjective, therefore making it more prone to individual worker bias as observed with the moderate agreement between workers in Tab.~\ref{tab:icc-values}. In \ref{SetupOne} the scores are skewed towards the left, indicating most workers found the system responses less interesting~(see Fig.~\ref{fig:intworker-dist-s2}) with a median score of 37 compared to 50 for \ref{SetupTwo}. 

Scores for \emph{explanation quality} are skewed towards the left, with a median score of 25 (\ref{SetupOne}) and 22 (\ref{SetupTwo}), as shown in Fig.~\ref{fig:EQworker-dist-s1} and \ref{fig:EQworker-dist-s2}. This shows that even though most workers find that system recommendations relevant there is a lack of explanation on why the recommendations are made. These findings are in line with recent work conducted by~\citet{explainableCRS-Shuyu}, showing that most \ac{CRS} dialogues lack explanation in their recommendations.

\dnegskip
\section{Effect of user feedback}
\label{section:RQ1}
\halfnegskip
In this section, we answer \ref{item:RQ1}: How does user feedback from the follow-up utterance influence the evaluation labels collected from both crowdworkers and \acp{LLM}?

\begin{figure}[!t]
    \centering
    \begin{subfigure}{0.4\columnwidth}
        \includegraphics[width=\linewidth]{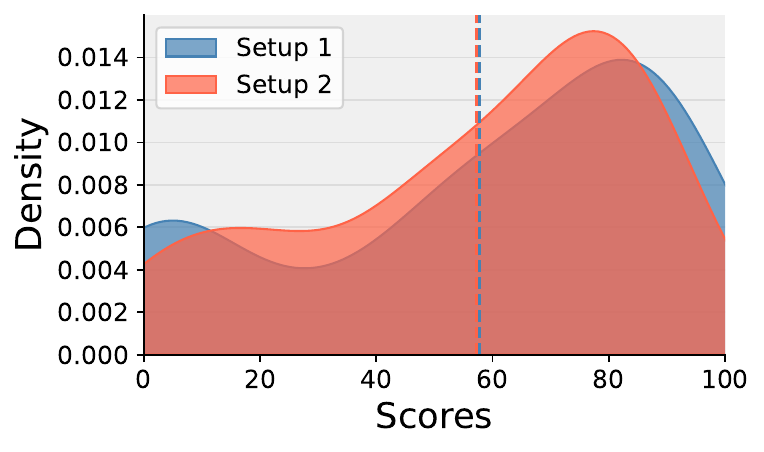}
        \vspace*{-6mm}
        \caption{Relevance}
        \label{fig:reldial}
    \end{subfigure}
    \quad
    \begin{subfigure}{0.4\columnwidth}
        \includegraphics[width=\linewidth]{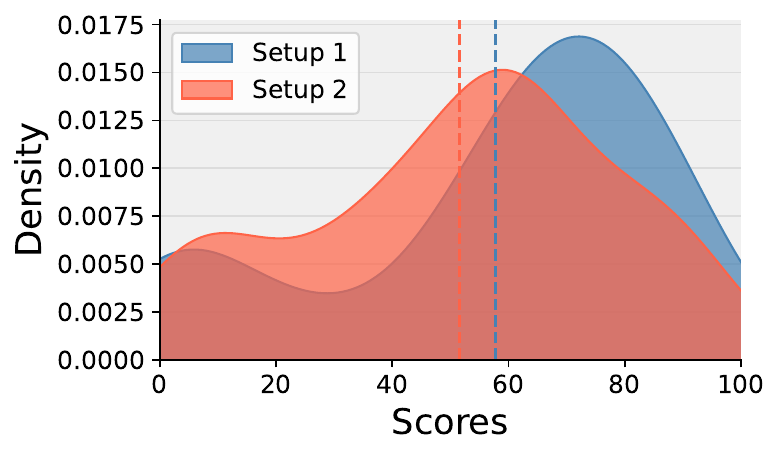}
        \vspace*{-6mm}
        \caption{Usefulness}
        \label{fig:usedial}
    \end{subfigure}
    \\
    \begin{subfigure}{0.4\columnwidth}
        \includegraphics[width=\linewidth]{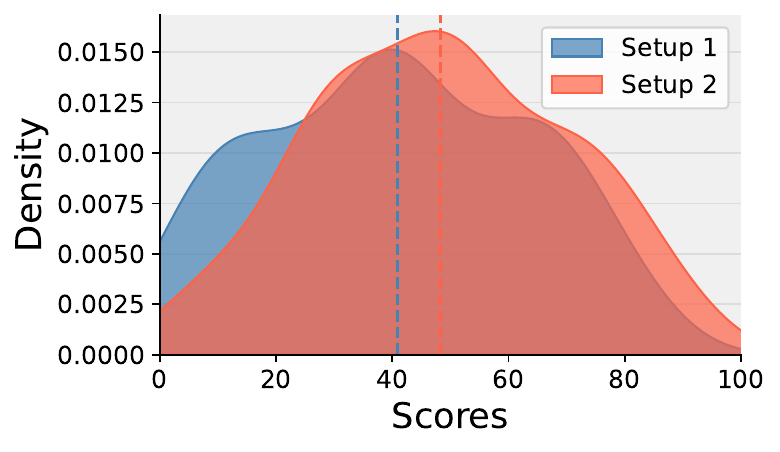}
        \vspace*{-6mm}
        \caption{Interestingness}
        \label{fig:intdial}
    \end{subfigure}
    \quad
    \begin{subfigure}{0.4\columnwidth}
        \includegraphics[width=\linewidth]{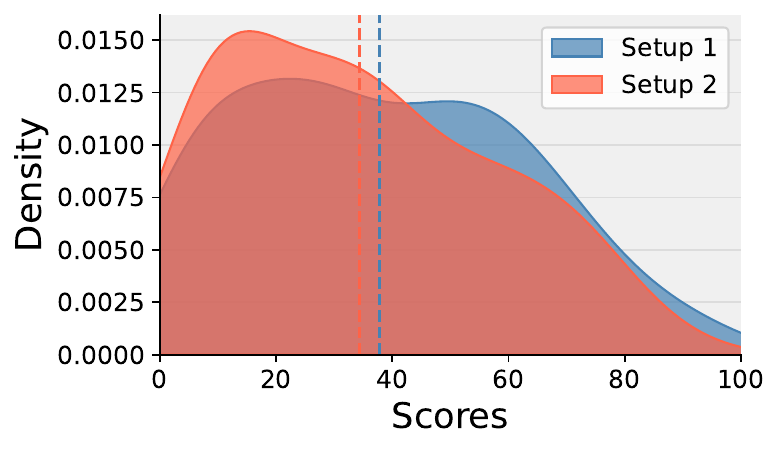}
        \vspace*{-6mm}
        \caption{Explanation quality}
        \label{fig:EQdial}
    \end{subfigure}
   \\
    \begin{subfigure}{0.4\columnwidth}
        \includegraphics[width=\linewidth]{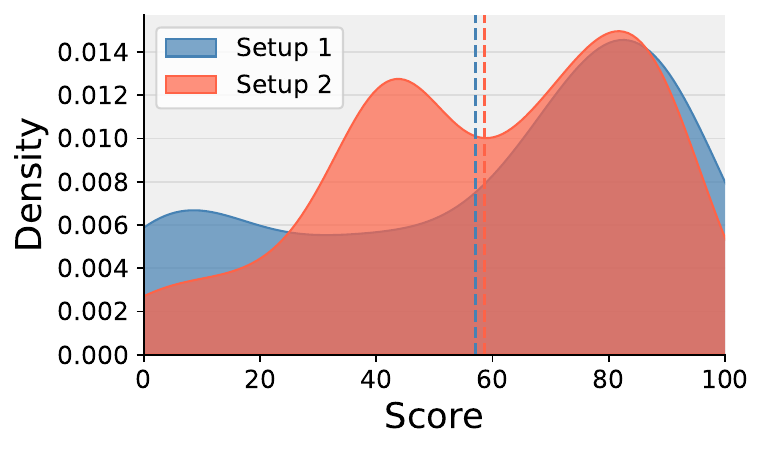}
        \vspace*{-6mm}
        \caption{Relevance-LLM}
        \label{fig:llm-dist-rel}
    \end{subfigure}
    \quad        
    \begin{subfigure}{0.4\columnwidth}
        \includegraphics[width=\linewidth]{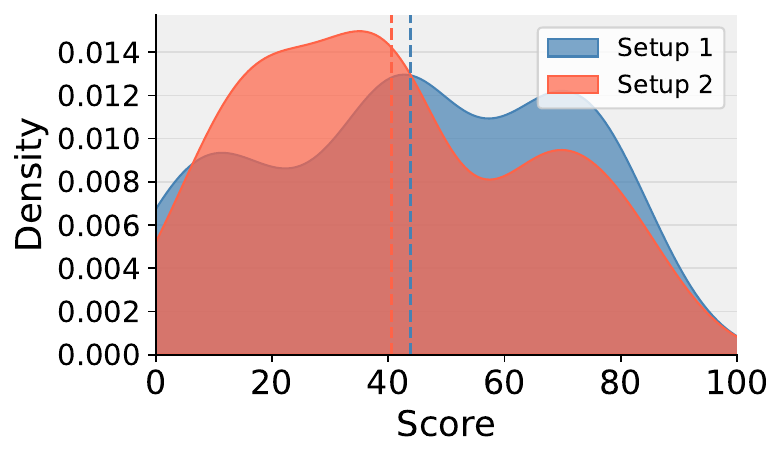}
        \vspace*{-6mm}
        \caption{Usefulness-LLM}
        \label{fig:llm-dist-use}
    \end{subfigure}
    \begin{subfigure}{0.4\columnwidth}
        \includegraphics[width=\linewidth]{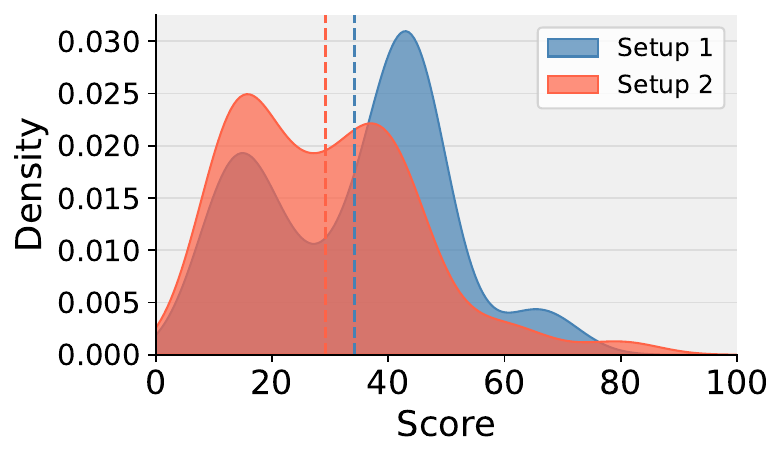}
        \vspace*{-6mm}
        \caption{Interestingness-LLM}   
        \label{fig:llm-dist-int}
    \end{subfigure}
    \quad
    \begin{subfigure}{0.4\columnwidth}
        \includegraphics[width=\linewidth]{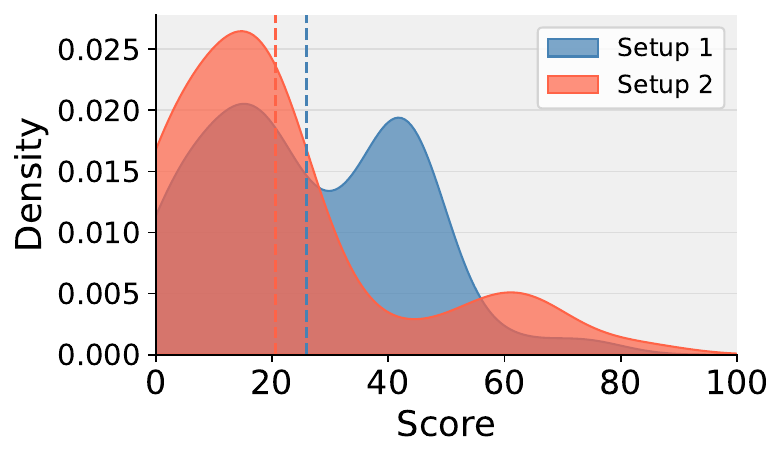}
        \vspace*{-6mm}
        \caption{Explanation qual.-LLM}   
        \label{fig:llm-dist-EQ}
    \end{subfigure} 
    \caption{\Acl{KDE} plots comparing aggregated crowdworker and LLM scores for both setups. The dotted lines represent the overall mean for each setup. }
    \label{fig:dialogue-distributions}
\end{figure}

\header{Distributions}
For the \textit{crowdsourced labels}, each turn is annotated by three workers; their ratings are averaged to get the overall score per turn.
Fig.~\ref{fig:dialogue-distributions}~(a)--(d) show the density distributions of the scores:
\textit{Relevance and usefulness} are skewed towards the right~(Fig.~\ref{fig:reldial} and \ref{fig:usedial}) for both setups, showing that more turns are found to be more relevant and useful by the crowdworkers. 
For \textit{usefulness}, the peak for \ref{SetupTwo} is towards the center compared to \ref{SetupOne}, indicating a decrease in the number of turns that are highly useful as workers have access to the user's follow-up utterance, adding more context during the assessment. 
Cases where the system makes a relevant recommendation, are rated highly useful in \ref{SetupOne}, but the user's feedback in \ref{SetupTwo} changes the worker's rating to lower values in certain cases. 
E.g., in cases where the user has already watched the movie or even though the movie satisfies their requirements (e.g., genre and actor), they do not like other aspects.
\textit{Interestingness} has a more central peak with a wide range showing there was a lot of variability in the assessment~(Fig.~\ref{fig:intdial}). More turns are assessed as interesting in \ref{SetupTwo} compared to \ref{SetupOne}, with fewer turns being scored as highly interesting in both setups~(80--100). 
Similar observations pertain to \textit{explanation quality}~(Fig.~\ref{fig:EQdial}). \looseness=-1

We also plot the distribution of scores from the \ac{LLM} in Fig.~\ref{fig:dialogue-distributions}~(e)--(f). The \textit{relevance} \ac{KDE} plot exhibits a dual-peak distribution with a minor peak at lower values and a more pronounced peak at higher values, notably around 80 and above~(Fig.~\ref{fig:llm-dist-rel}). In \ref{SetupTwo}, the \ac{KDE} plot shows a distribution peaking in the mid to higher range of the score scale.

In contrast, \textit{usefulness} (Fig.~\ref{fig:llm-dist-use}) shows three peaks in \ref{SetupOne} (not significant from each other), with \ref{SetupTwo} having a distinctive peak between scores of 10--40.  The slight peak towards the high scores compared to \ref{SetupOne} indicates that with the user's follow-up utterance, the \ac{LLM} finds most turns not to be highly useful.
We observe a different pattern for \textit{interestingness} and \textit{explanation quality} with scores skewed towards the left showing that the \ac{LLM} rates most turns low on interestingness and explanation similar to observations made from the crowdworker scores.

\vspace*{0.1mm}
\header{Humans vs.\ LLM} The different peaks between the two setups across the four aspects indicate a significant divergence in how crowdworkers and \acp{LLM} perceive and rate the aspects in different setups. \ref{SetupOne} is characterized by high peaks in high-range scores, compared to \ref{SetupTwo} which exhibits peaks in the moderate range except for relevance, which has both moderate and high peaks. This contrast suggests that user feedback from the follow-up utterance has a notable impact on both the crowdworkers and \ac{LLM} assessments.

\begin{figure}[!htb]
    \centering
    \begin{subfigure}{0.4\columnwidth}
        \includegraphics[width=\linewidth]{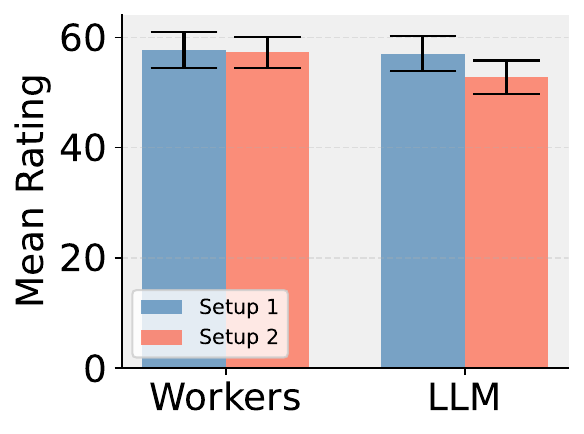}
        \vspace*{-6mm}
        \caption{Relevance}
        \label{fig:relmean}
    \end{subfigure}
    \quad
    \begin{subfigure}{0.4\columnwidth}
        \includegraphics[width=\linewidth]{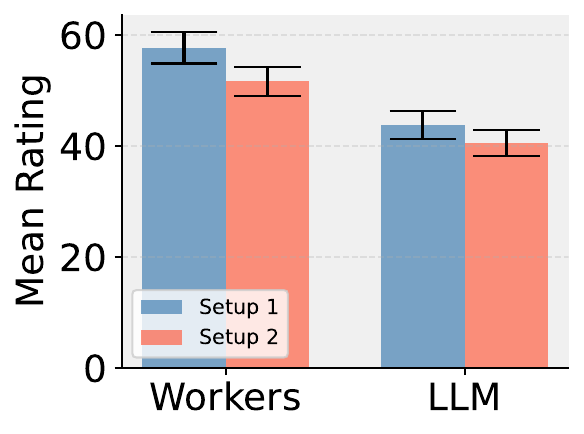}
        \vspace*{-6mm}
        \caption{Usefulness}
        \label{fig:usemean}
    \end{subfigure}
    \\
    \begin{subfigure}{0.4\columnwidth}
        \includegraphics[width=\linewidth]{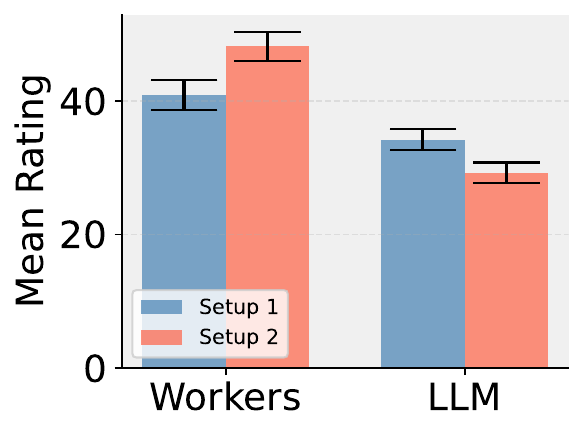}
        \vspace*{-6mm}
        \caption{Interestigness}
        \label{fig:intmean}
    \end{subfigure}
    \quad
    \begin{subfigure}{0.4\columnwidth}
        \includegraphics[width=\linewidth]{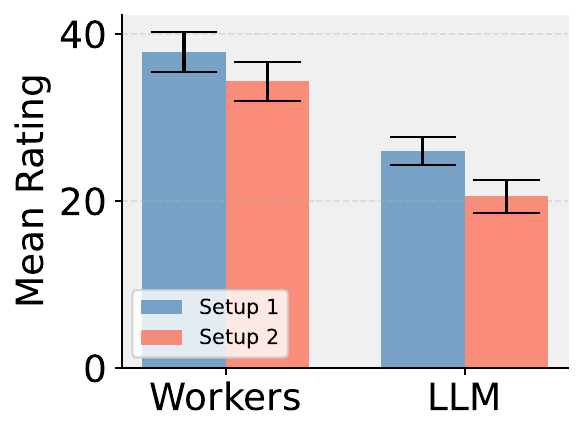}
        \vspace*{-6mm}
        \caption{Explanation quality}
        \label{fig:EQmean}
    \end{subfigure}    
    \caption{Mean rating for each aspect across the two setups, for both the crowdworkers and LLM. }
    \label{fig:mean-rating}
\end{figure}

\header{External agreement} 
Next, we compute the overall mean score for each setup for both the crowdworkers and the LLM with confidence intervals; see Fig.~\ref{fig:mean-rating}. 
There is no significant difference in \textit{relevance scores} between the two annotator groups. This indicates that the presence of the user's follow-up utterance does not significantly affect the relevance assessment. Relevance primarily relies on the system's ability to provide a topically relevant recommendation to the user's request, and having only the user's initial request appears sufficient for this assessment. 
However, some differences in relevance scoring emerge when the follow-up utterance is available to certain workers. 
In many instances, workers influenced by the follow-up utterances tend to assign high scores to non-relevant system responses if the user accepts the recommendation, even if it deviates from their initial request. 
Conversely, they may assign low scores to relevant system responses if the user dislikes the recommendation based on specific attributes, despite its topical relevance. 
Similar patterns are observed in the scores assigned by \ac{LLM}, with the \ac{LLM} having a low mean score for \ref{SetupTwo}.

\ac{LLM}-based annotations are consistently lower in terms of \textit{usefulness scores} compared to crowdworkers', as indicated by lower mean scores in both setups~(Fig.~\ref{fig:usemean}). The mean scores are statistically significant for crowdworkers but not for the \ac{LLM}, highlighting the substantial influence of the follow-up utterance on usefulness assessments.
In \ref{SetupOne}, crowdworkers assign high usefulness scores to system responses, closely aligning with relevance scores. This suggests that annotators assess usefulness like relevance in the absence of the follow-up utterance.

Conversely, in \ref{SetupTwo}, there is a significant drop in the mean usefulness score. This reflects workers transitioning from assessing relevance to considering how well the system response addresses various facets of the user's needs, often revealed in the follow-up utterance. 
E.g., a user may request an action movie initially, but specific preferences may emerge in the subsequent utterance, such as actor or director preferences as users typically reveal their complete information needs through a back-and-forth exchange~\citep{AliannejadiZCC19}.

In contrast to other aspects, \textit{interestingness} presents an intriguing observation. Workers assign lower scores in \ref{SetupOne} compared to \ref{SetupTwo}~(Fig.~\ref{fig:intmean}). 
Both groups of annotators assign lower scores for interestingness in both setups, and these differences in mean scores are statistically significant. 
Examination of annotators' justifications reveals that they hold strict criteria for rating system responses as interesting and had relatively high expectations for a response to be deemed interesting. 
This is reflected in the score distributions depicted in Fig.~\ref{fig:intdial}, where a smaller proportion of turns receive a score of 100 for interestingness.
The disparity between the setups is particularly notable in \ref{SetupOne}, where only 2\% of the turns received a score of 90 or higher, as opposed to 7\% in \ref{SetupTwo}.

The mean score for explanation quality is not statistically significant for crowdworkers, although there is a noticeable drop of over 3 points from \ref{SetupOne} to \ref{SetupTwo}. However, it is statistically significant for \ac{LLM} annotators. It is worth noting that this aspect consistently yields low mean scores compared to the other aspects, ranging from 20 to 37, as indicated by the mean bars in Fig.~\ref{fig:EQmean}. Fewer turns receive a perfect score of 100 in the aggregated scores for \ref{SetupTwo}; see Fig.~\ref{fig:EQdial}. Both annotator groups agree in assigning lower scores for this aspect, highlighting the lack of recommendation explanations in the dataset. An analysis of annotator justifications reveals varying expectations regarding system explainability. While some workers expect the system to provide explanations related to aspects like the movie's cast or director, others focus on different facets. This subjectivity in worker bias and expectations regarding system explanations contributes to the variation in scores for explanation quality.

\header{Humans vs.\ LLM}
Overall, both groups agree on assessing relevance, but differences emerge when considering follow-up utterances, influencing relevance scores for \ac{LLM}. \acp{LLM} consistently assign lower usefulness scores than human annotators, indicating the challenge of defining usefulness when follow-up utterances reveal complex user needs. Unlike humans, \acp{LLM} do not personalize the system's usefulness to the user. This highlights the importance of including follow-up utterances for more accurate evaluation labels that reflect the user's perspective. Both groups agree on the lack of explanations from the system.

\header{Agreement with expert ratings} Here, we examine how well human and \ac{LLM} labels align with expert ratings. We collect expert ratings on the user satisfaction aspect to investigate the correlation of the fine-grained aspects to overall user satisfaction following~\citet{Siro-UsatCRS}. Using the same setup, we collect the expert ratings from two experts. Since our initial annotation was on an S100 scale we transformed the labels to S3 scale~(1--3) for all aspects~\citep{transfroming-relevance-scale} and then calculated the Spearman's $r$ between each aspect and the expert rating. We report our results in Tab.~\ref{tab:expertcorr}. 

\begin{table}[!t]
    \centering
    \caption{Spearman's $r$ correlation coefficient between the aspects and expert user satisfaction ratings for both the crowdworkers and \ac{LLM}. * indicate non-significant values ($p< 0.05$).}
    \begin{tabular}{l cc cc}
    \toprule
         & \multicolumn{2}{c}{Crowdworkers} & \multicolumn{2}{c}{LLM} \\
        \cmidrule(r){2-3}
     \cmidrule(r){4-5}
     Aspects & \ref{SetupOne} & \ref{SetupTwo}& \ref{SetupOne} & \ref{SetupTwo} \\
        \midrule
        Relevance & 0.56  & 0.51 & 0.63 & 0.52  \\
        Usefulness & 0.55 & 0.66 & 0.45 & 0.41 \\
        Interestingness & 0.31\rlap{*} & 0.39 & 0.27\rlap{*} & 0.21\rlap{*} \\
        Explanation quality & 0.44 & 0.42 & 0. 54 & 0.50 \\
        \bottomrule
    \end{tabular}    
    \label{tab:expertcorr}
\end{table}

In \ref{SetupOne}, relevance displays a moderate positive correlation of 0.56 (crowdworkers) and 0.63 (\ac{LLM}) with expert ratings, indicating a similar alignment with expert satisfaction in the absence of follow-up utterances. Usefulness shows stronger correlations, with 0.55 for crowdworkers and 0.45 for \ac{LLM}, suggesting that usefulness judgments are significantly influenced by the absence of follow-up utterances in this setup. Interestingness exhibits weaker correlations in both groups, suggesting potential challenges or subjectivity in assessing this aspect. Explanation quality demonstrates moderate correlations (0.47 for crowdworkers and 0.54 for \ac{LLM}), indicating moderate alignment with expert satisfaction ratings.

In \ref{SetupTwo}, relevance maintains positive correlations, 0.51 (crowdworkers) and 0.52 (\ac{LLM}). Usefulness shows notably stronger correlations, 0.66 for crowdworkers and  0.41 for \ac{LLM}, indicating that crowdworkers assign scores closely aligned with the user feedback. Interestingness continues to exhibit weaker correlations (0.39 for crowdworkers and 0.21 for \ac{LLM}). Explanation quality, while still aligned with expert ratings, has slightly weaker correlations (0.44 for crowdworkers and 0.50 for \ac{LLM}) compared to \ref{SetupOne}.

\header{Humans vs.\ LLM} Correlations with expert ratings highlight that relevance and usefulness assessments generally have a stronger alignment with expert user satisfaction ratings across setups and annotator types. However, interestingness shows weaker correlations, indicating potential challenges in assessing this aspect consistently and objectively~\citep{Siro-satTDS}. 
In general, we note that humans perform well in assessing user experience measures such as usefulness and interestingness while \acp{LLM} perform well in assessing utility measures such as relevance and explanation quality.

Further analysis of the correlation of the aspects in the two setups and task duration is available in our \nameref{sec:appendix}.

\negskip
\section{Significance of User Feedback} 
\label{section:RQ2}
\halfnegskip
In this section, we examine the impact of user feedback from follow-up utterances on reducing annotator variability in their evaluation labels as part of our analysis to answer \ref{item:RQ2}. To assess agreement, we calculate the standard deviation of workers' scores for each turn and categorize the data into two groups: \textbf{Group~1} (scores below the median standard deviation, indicating high agreement) and \textbf{Group~2} (scores above the median, indicating low agreement). We find that \textit{interestingness} and \textit{explanation quality} consistently exhibit higher agreement among annotators when the system response is uninteresting or lacks explanation. However, there is no clear agreement pattern among workers for \textit{relevance} and \textit{usefulness}.

\begin{figure}[!htb]
    \centering
    \begin{subfigure}{0.4\columnwidth}
        \includegraphics[width=\linewidth]{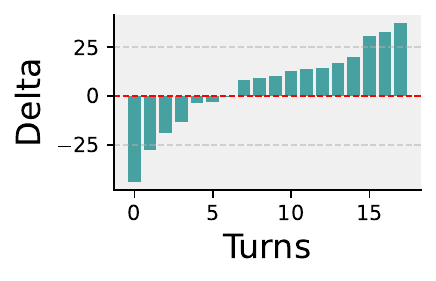}
        \vspace*{-6mm}
        \caption{Relevance}
        \label{fig:deltagrouprel}
    \end{subfigure}
    \quad
    \begin{subfigure}{0.4\columnwidth}
        \includegraphics[width=\linewidth]{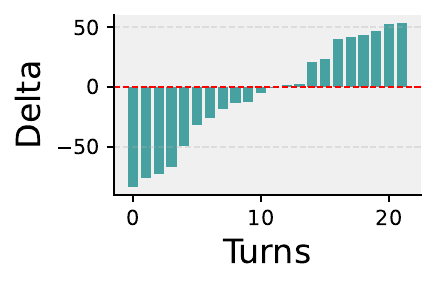}
        \vspace*{-6mm}
        \caption{Usefulness}
        \label{fig:deltagroupuse}
    \end{subfigure}
    \\    
    \begin{subfigure}{0.4\columnwidth}
        \includegraphics[width=\linewidth]{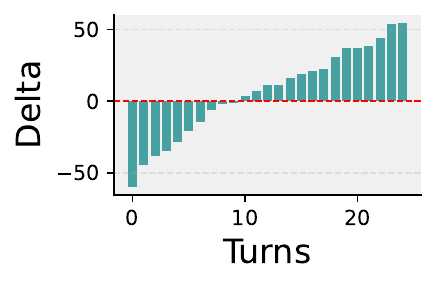}
        \vspace*{-6mm}
        \caption{Interestingness}
        \label{fig:deltagroupint}
    \end{subfigure}
    \quad
    \begin{subfigure}{0.4\columnwidth}
        \includegraphics[width=\linewidth]{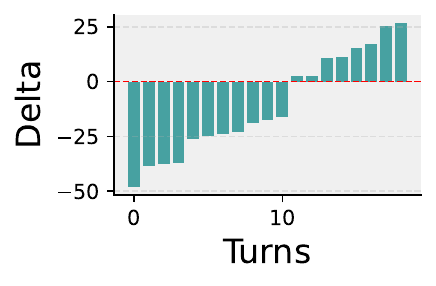}
        \vspace*{-6mm}
        \caption{Explanation quality}
        \label{fig:deltagroupEQ}
    \end{subfigure}
    \caption{Difference in scores assigned to dialogues turns for four aspects in Group~1 with low variability vs.\ dialogues in Group~2 with high variability between the worker scores from the mean rating. %
    }
    \label{fig:deltagroup}
\end{figure}

We compare turns in Group~2 from \ref{SetupOne} with Group~1 from \ref{SetupTwo}, where Group~2 initially exhibits high variability in evaluation labels, but shows increased agreement in \ref{SetupTwo} due to the presence of the user's follow-up utterance. Specifically, there are 18 turns for relevance, 22 for usefulness, 25 for interestingness, and 19 for explanation quality in this analysis (Group~2 initially consisted of 48 to 50 turns). Overall, at least 30\% of the turns demonstrate improved worker agreement in \ref{SetupTwo}. To quantify score differences between Group~1 and Group~2, we calculate their delta and present the results in Fig.~\ref{fig:deltagroup}. We observe significant score differences for the same instances under different conditions. Relevance and interestingness have more turns rated highly in Group~1 (positive delta scores)~(Fig.~\ref{fig:deltagrouprel} \& \ref{fig:deltagroupint}) while usefulness and explanation quality have turns rated low in Group~1~(Fig.~\ref{fig:deltagroupuse} \& \ref{fig:deltagroupEQ}).

\header{Manual analysis}
Our manual analysis primarily focused on the usefulness aspect due to its substantial impact, with the highest mean delta difference (35) compared to other aspects (interestingness: 26, explanation quality: 22, relevance: 15). We analyze 22 turns to identify instances where the user's follow-up utterances notably enhance worker agreement, shedding light on cases where the presence of the user's next utterance significantly improves consensus. \looseness=-1

This analysis identifies specific scenarios where user feedback plays a pivotal role, such as addressing ambiguous requests by providing clarity, making generic requests more specific and actionable, simplifying complex requests, and compensating for annotators' lack of domain knowledge. In these scenarios, user feedback consistently improves the overall quality and consistency of the annotation process, highlighting its significance in enhancing system evaluations.\footnote{A detailed example of a complex user request can be found in our \nameref{sec:appendix} in Sec.~B.1. %
}

Apart from resolving uncertainty in user requests, follow-up utterances are crucial when annotators encounter unfamiliar topics. An analysis of annotators' justifications reveals that when annotators lack prior knowledge, the user's knowledge about the recommended item, coupled with explicit feedback, bridges the knowledge gap, resulting in more precise evaluations of the system's performance, even when annotators lack subject matter expertise. \looseness=-1

\section{Sources and Bias} 
\label{section:RQ3}
\halfnegskip
In this section, we answer~\ref{item:RQ3}. To understand the basis of workers' assessments and their choices when assigning evaluation labels, we conducted a survey where we asked workers to indicate the sources of information they relied on when making judgments. Both setups offered the same options for information sources, except for follow-up utterance, which was available only in \ref{SetupTwo}. The available sources included \textit{personal knowledge, searched online, guessed, user request,} and \textit{system response}. 
Additionally, we performed a manual analysis of the workers' justifications to evaluate any potential biases introduced by their chosen information sources.

\begin{figure}[!hbt]
    \centering
    \begin{subfigure}{0.4\columnwidth}
        \includegraphics[width=\linewidth]{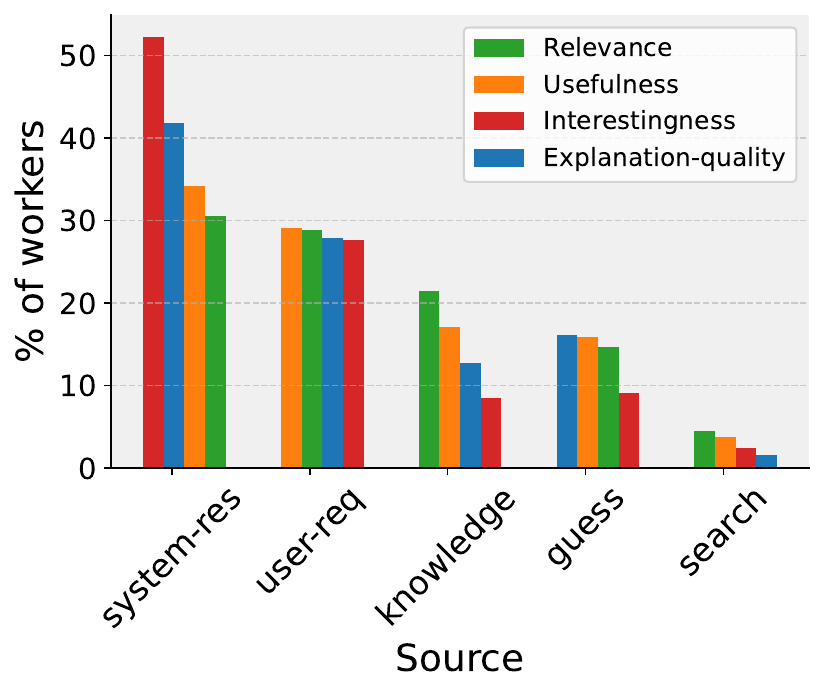}
        \vspace*{-6mm}
        \caption{\ref{SetupOne}}
        \label{fig:sources-without}
    \end{subfigure}
    \quad
    \begin{subfigure}{0.4\columnwidth}
        \includegraphics[width=\linewidth]{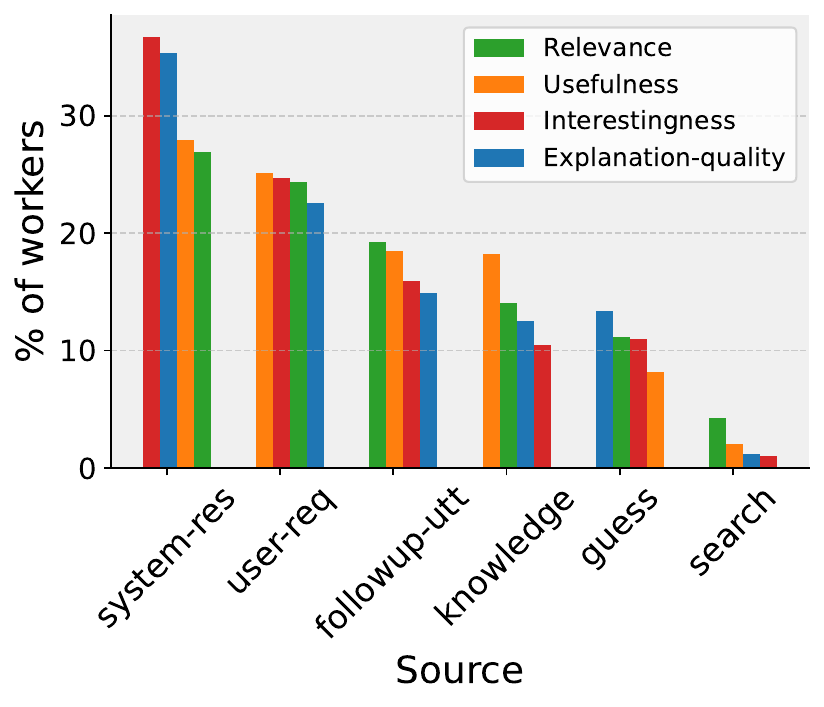}
        \vspace*{-6mm}
        \caption{\ref{SetupTwo}}
        \label{fig:sources-with}
    \end{subfigure}
    \caption{Distribution of sources workers relied on to make their judgments for the four aspects in the two setups.}
    \label{fig:sources}
    \vspace{-2mm}
\end{figure}

\header{Sources} \label{sources}
Fig.~\ref{fig:sources} shows the percentage of workers who relied on each information source during evaluation. The x-axis represents the information sources, while the y-axis indicates the percentage of workers relying on each source. Across both setups, it is evident that workers predominantly depend on information within the dialogue itself, specifically the user's request and the system's response, for their assessments. Interestingly, the evaluation of interestingness and explanation quality was primarily influenced by the system's response. While explanation quality considered both the user request and system response, we observe a notably higher reliance on the system's response compared to the user request.
On the contrary for relevance and usefulness assessment, we observe that workers mostly rely on the user request to ensure the system meets the user's need. However, there is a marginal difference between system response and user request for these two aspects showing that the two are equally important during assessment. 
Without the user's follow-up utterance, some workers make an educated guess on the relevance and usefulness of the system response.

In \ref{SetupTwo}~(Fig.~\ref{fig:sources-with}), we note a drop in the percentage of workers relying on the system response during evaluation, showing that the follow-up utterance introduces another dynamic to be considered by the workers during the assessment. Usefulness which measures how well the system response meets the user's needs has a high percentage of workers relying on the user's follow-up utterance. Unexpectedly we note a high number where workers relied on personal knowledge to gauge the usefulness of the system response, showing that they introduce personal bias in assessing this aspect.

A few workers utilize online sources, primarily when assessing relevance and usefulness, implying that workers without domain knowledge leverage online information for more accurate assessments. Interestingly, there is a decrease in the percentage of workers using online sources in \ref{SetupTwo}, specifically for evaluating usefulness. This suggests that the introduction of the user's follow-up utterance in \ref{SetupTwo} acts as an additional information source, assisting workers lacking domain knowledge in assessing the system response's usefulness.

\header{Biases}
Several studies highlight the influence of biases on crowdworkers' judgments~\citep[e.g.,][]{mitigating-workerbias,DBLP:cognitiveload-ds,Carsten-cognitive-bias}. In our work, we specifically explore how the sources of information outlined in Sec.~\ref{sources} introduce biases into crowdsourced labels.  Fig.~\ref{fig:sources} illustrates a reliance on online sources for assessment, which, while potentially augmenting workers' domain knowledge, can introduce specific biases, such as popularity bias~\citep{popularitybias-survey}. From the analysis of workers' justifications, we observe instances in assessing usefulness where some workers forego the user's feedback on the system's response and rely on online movie reviews. Justifications like ``The movie seems to be liked by many so it is useful,'' and ``The movie is not rated highly'' are observed, indicating that these external sources could bias workers, leading to ratings that may not accurately reflect the user feedback.

A notable percentage of workers rely on the user's request when assessing interestingness. However, interestingness assessments should primarily be based on the system's response. 
Therefore, we note several workers get biased by the relevance of the recommended item to the user request during their assessments. To mitigate this bias, it may be prudent to restrict access to the user request when evaluating aspects like interestingness, where the focus is on assessing the system independently of the user's input. \looseness=-1

In comparison to other aspects, we notice that workers rely on their knowledge to evaluate the usefulness of the system response~(see Fig.~\ref{fig:sources}), introducing personal preference bias despite the explicit tie to the user's needs. Also, workers display a bias towards rating longer system responses highly for explanation quality.

\dnegskip
\section{Discussion and Conclusion}
\halfnegskip
In this work, we addressed the question of how the inclusion of user feedback, both implicit and explicit, from the user's follow-up utterance, influences the evaluation labels from crowdworkers and \acp{LLM}. 
Our analysis revealed intriguing patterns across various aspects, providing valuable insights into the impact of user feedback on the quality of assessments.

\textbf{Relevance.} We considered two experimental setups, one without the user's follow-up utterance and one with.
In both setups, crowdworkers and \ac{LLM}-based annotators largely concur when evaluating relevance. 
However, subtle differences emerge with the inclusion of follow-up utterances, particularly for \acp{LLM} which tend to assign lower scores compared to humans. 
Though there is no significant mean difference in \ref{SetupOne} for both annotator groups, we note that \ac{LLM} scores show a higher correlation to expert user satisfaction ratings than humans. 
Crowdworkers relied more on the system response, user request, and their prior knowledge to gauge the relevance of the recommended item. 
With lots of candidate movies to recommend, crowdworkers may lack knowledge of some of these movies to assess their relevance, which results in making an educated guess. 
However, compared to humans \acp{LLM} are rich in internal knowledge on these movies, thus improving their relevance assessment. 

\textbf{Usefulness.} The usefulness ratings show a distinctive contrast between the two annotator groups. 
Human annotators displayed strong correlations with user satisfaction in \ref{SetupTwo}, suggesting their ability to personalize the system's usefulness to individual users. 
In contrast, \acp{LLM} consistently assigned lower usefulness scores in both setups, highlighting the challenge of assessing usefulness when follow-up utterances reveal conflicting user needs from their initial request. 
This shows that when user feedback conflicts with a model's internal knowledge it leads to inconsistency in the ratings. 

\textbf{Interestingness.} The aspect of interestingness presented unique challenges. 
Both crowdworkers and \acp{LLM} exhibited lower correlations with user satisfaction ratings, indicating that both annotator groups struggled to capture the user's subjective perception of interestingness. 
The presence of user feedback had a limited impact on improving assessments in this aspect. This is also observed with the low Kappa and \ac{ICC} scores in Tab.~\ref{tab:icc-values}. Nonetheless, utterances such as ``that's interesting,'' ``sounds good,'' and ``you are funny'' lead to an increase in annotator agreement and correlation with user satisfaction ratings~(\ref{SetupTwo}), emphasizing the significance of user feedback in improving system evaluation.

\textbf{Explanation quality.} Both annotator groups concur on the absence of explanations provided by the system. 
This shared observation underscores a significant limitation in current \ac{CRS}, as both human evaluators and \acp{LLM} noted the lack of explanatory content in system responses. 
The \ac{LLM} shows less sensitivity to user feedback with high correlating scores to overall user satisfaction compared to humans. 
Humans' ratings are affected by their personal expectations of the system's explainability, which is not evident in the \ac{LLM} scores. 
This shows that \acp{LLM} can maintain objectivity when assessing system performance.

In general, there is a distinct difference in ratings assigned by both annotator groups in the two setups, indicating that user feedback does influence system evaluation. 
Human workers are more susceptible to user feedback in usefulness and interestingness compared to \acp{LLM} in interestingness. 
User feedback leads to personalized usefulness assessment by workers and improves worker agreement when uncertainty arises in the user's request. 
The lack of adaptability to user feedback by the \ac{LLM} in assessing usefulness, suggests that \acp{LLM} may require additional mechanisms such as prompt tuning to enhance user-centric evaluations by \acp{LLM}. Therefore, it is important to assign annotation tasks to LLMs based on the nature of the task, leveraging their strengths in objective assessments like relevance annotation while complementing with human assessors for tasks demanding subjective evaluation or sensitivity to user preferences and feedback. Combining human annotators and \acp{LLM} can lead to better system evaluations by leveraging the unique strengths of each type of annotator for specific evaluation tasks. \looseness=-1

However, user feedback can sometimes lead to assessments that do not align with overall user satisfaction, resulting in lower correlation scores as observed for the relevance aspect in \ref{SetupTwo}. 
It is important to note that this study employed a single \ac{LLM} for annotation, and results may vary with different \acp{LLM}. 
Additionally, potential biases in the crowdworker pool and the LLM's training data could influence the findings. For future work, we will conduct further research to validate our findings across diverse conversational systems.

\subsubsection*{\bf Online Appendix}
\label{sec:appendix}
Further analyses and discussions are available at \url{https://github.com/Clemenciah/LLMCrowdDialogueEval/tree/main/Appendix}.

\subsubsection*{\bf Data} We
publicly release the annotated data at \url{https://github.com/Clemenciah/LLMCrowdDialogueEval/tree/main/Data}.

\subsubsection*{\bf Acknowledgments}
This research was supported by the Dreams Lab, a collaboration between Huawei Finland, the University of Amsterdam, and the Vrije Universiteit Amsterdam, 
by 
the Dutch Research Council (NWO), under project numbers 024.004.\-022, NWA.\-1389.20.\-183, and KICH3.LTP.20.\-006, and 
by the European Union's Horizon Europe program under grant agreement No 101070212.

All content represents the opinion of the authors, which is not necessarily shared or endorsed by their respective employers and/or sponsors.

\balance
\bibliographystyle{ACM-Reference-Format}
\bibliography{references}

\end{document}